\documentstyle[12pt]{article}

\def\applt{\;\; {\lower3pt\hbox{$
{\buildrel < \over {\scriptstyle \sim} }$}}\;\;}

\large
\begin{document}
\newcommand{\cl}{\centerline}
\newcommand{\beq}{\begin{equation}}
\newcommand{\eeq}{\end{equation}}
\newcommand{\beqa}{\begin{eqnarray}}
\newcommand{\eeqa}{\end{eqnarray}}
\def\d{{\rm d}}
%%%%%%%%%%%%%%%%%%%%%%%%
\hbox{\hskip 4.0 true in ITP-US-93-2}
\vbox{\vskip 0.3 true in}
 \centerline{\large{{\bf Power Corrections to QCD Sum Rules  }}}\par
\vbox {\vskip 0.1 true in}
\centerline{\large{{\bf  for Compton Scattering.}}}\par

\vbox {\vskip 0.3 true in}

%\centerline{\bf , June 21, 1993}
\centerline{Claudio Corian\`{o}}
\centerline{ Institute for Theoretical Physics}
\centerline{ University of Stockholm}
\centerline{Box 6730, 113 85 Stockholm, Sweden}
\centerline{and}
\centerline{Institute For Theoretical Physics}
\centerline{State University of New York at Stony Brook}
\centerline{Stony Brook, 11794 NY, USA}
\vbox {\vskip 0.2 true in}

\centerline {ABSTRACT}
\medskip
\narrower{We extend previous work
on sum rules for the invariant amplitudes of pion Compton scattering
 by deriving
a complete lowest order perturbative
spectral function  - and its leading non perturbative power corrections -
 for a specific combination of the two helicities $(H_1 + H_2)$ of this
process.
Using some properties of a modified version of the Borel transform,
we develop a method of calculation of the  gluonic corrections
which can be easily extended to  other similar reactions, such as proton
Compton
scattering.
A preliminary comparison of the new sum rule with the  pion form factor sum
rule
is made.

\vbox{\vskip 1.0 true in}
\noindent{\footnotesize email: claudio@max.physics.sunysb.edu} \\
%\noindent $^{(*)}$ {\footnotesize present address}

\newpage
%\cl{Institute for Theoretical Physics}\par
%\cl{University of Stockholm}\par
%\cl{S 113 Vanadisvagen, Stockholm, Sweden}\par
%\cl{and}
%\cl{Istituto di Fisica}\par
%\cl{Universita' degli Studi di Parma}
%\cl{INFN Gruppo Collegato di Parma, Italy}

\section{Introduction}

Among the semi-phenomenological methods developed  in the past in
the description of  the resonant region of QCD,
sum rules are
the closest to a fundamental description
of the strong interactions and they have been formulated  within a
calculational
scheme which is very similar to perturbative QCD. In the sum rules approach a
"master equation"   relates the (timelike) region of the external invariants
in the correlation functions, the region where the resonances are located,
to the euclidean region where
information from the operator product expansion is parameterized by local
vacuum condensates (power corrections). In practical applications, only the
phenomenological input from the lowest dimensional condensates (quark and gluon
condensates) is required while the resonance region is
described
by a model spectral density of states which contains non perturbative
information on the
lowest resonance, although mixed with the continuum contribution.

A particular class of sum rules are {\em Borel sum rules}. They are obtained
by acting with Borel transforms on the master equation, in order to enhance the
resonant contribution respect to the contribution from the higher states, while
a
superconvergence condition on the same spectral density allows to remove the
continuum contribution  beyond a given "duality interval" ($s_0$).
 The net
result of this procedure is a "sum rule" which relates properties of the lowest
resonance to
an operator product expansion (ope) series which includes both perturbative and
non perturbative terms.
This way of obtaining information about the resonances, which works
very well for both 2 and 3 point functions, has largely contributed to our
understanding
of QCD in the intermediate energy region \cite{SVZ,NR1,JS}.
In the extension of the method from 3 to 4 point functions
various new difficulties are encountered,
such as  1) a more complex dependence of the dispersion relations on both
the invariants $s$ and $t$,
and 2) the question related to the validity of the ope.
As discussed in Ref. \cite{CRS}, this last fact
 constraints the size of the kinematical  $s$ and $t$
region in which the ope is applicable, demanding to consider by this method
scattering at fixed angle
($-t/s$ fixed). Once these kinematical conditions are satisfied, the {\em new }
information
avaliable  from sum rules for 4-point functions, compared to 3 point functions,
comes from the more general $s$ and $t$ dependence of the residui
at the double poles of the lowest resonance (the invariant amplitudes),
which can then be related to a perturbative spectral function and to its power
corrections.
Therefore, compared to form factors \cite{NR1,JS}, in the case of 4-point
functions
a true angular dependence of the sum rule comes into
play and a parallel study of the result
obtained by this method and by the factorization picture, for Compton
scattering,
becomes possible.
This last aspect has
been preliminarly discussed in Ref. \cite{CL}, where predictions from the
asymptotic
local duality sum rule of Ref. \cite{CRS} and  a modified factorization formula
 which resums Sudakov effects have been compared.

 In the first part of this work (sections 2 and 3) we give a complete
- i.e. non asymptotic - derivation of the local
duality sum rule for pion Compton scattering, extending
the results obtained in Ref. \cite{CRS} and we show that
the local duality sum rule  can be organized in
a  rather compact form in the Breit frame of the pion, where the spectral
densities
become polynomial. For this purpose we use the cutting rules for the
propagators,
as discussed in \cite{CRS}.
 In sections 4 we use a modified Borel transform, first introduced in
 Ref. \cite{CRS}, to evaluate the gluonic power corrections to the spectral
 function. An original scheme of calculation of such contributions is
discussed.
 A more technical discussion of its implementation in the
 calculation of a specific coefficient
 of the expansion is briefly described in appendix B.
 In section 5 we discuss the evaluation of the power corrections proportional
 to the quark condensates and give the complete expression of the spectral
funtion
 which appears in the sum rule for the sum $(H_1 + H_2)$ of the two helicity
 amplitudes of Compton scattering.
In sections 6 the new Borel sum rule is derived and compared to the
corresponding one for the form
factor case, while our conclusions are in section 7.
 Our convention can be found in appendix A; section B briefly illustrates the
evaluation of a specific gluonic correction.

\section{Overview}
We start our discussion by first quoting the main results of Ref. \cite{CRS}.
As a starting point we consider the following four-point amplitude
\begin{eqnarray}
\Gamma_{\sigma\mu\nu\lambda}(p_1^2,p_2^2,s,t)&=&i\int{\rm d}^4x\,{\rm d}^4y
\,{\rm d}^4z\exp
(-ip_1\cdot x+ip_2\cdot y-iq_1\cdot z)
\nonumber \\
& &\times \langle 0|T\left(\eta_{\sigma}(y)J_{\mu}(z)J_{\nu}(0)
\eta_{\lambda}^{\dagger}(x)\right)|0\rangle \; ,
\label{tp}
\end{eqnarray}
where
\begin{eqnarray}
J_{\mu}=2/3\bar{u}\gamma_{\mu}u -1/3\bar{d}\gamma_{\mu}d,
\;\;\;\;\;\;
\eta_{\sigma}=
\bar{u}\gamma_5\gamma_{\alpha}d
\label{jd}
\end{eqnarray}
are the electromagnetic and
axial currents respectively of up and down quarks.
The choice of this specific correlator is motivated by the particular process
that we have in mind, namely Compton scattering of a pion.
 The pions lines of momenta $p_1$ and $p_2$ are assumed to be off-shell
 while the photons lines $(q_1,q_2)$ are on-shell.

The perturbative expansion of the correlator in eq.~(1) is represented by the
diagrams
in Figs.~1-3. Specifically, Fig.~1 represents the perturbative contribution
to the expansion,  while
Figs.~2 and 3 describe the power corrections
which are proportional to the condensates of quarks (Figs.~2) and
gluons (Figs.~3). They parametrize the non
perturbative structure of the QCD vacuum.
As discussed in Ref. \cite{CRS}, in the case of pion Compton scattering, when
the
two external photons are physical and all the Mandelstam invariants
are moderately large (see also the discussion in Ref. \cite{CL}),
the contribution to the leading scalar
spectral function comes from the diagram in Fig. 5a.
This diagram differs from the corresponding one in the form factor case
(Fig. 5b) for having one additional internal line (the top line connecting the
two photons, see Fig.~5a) off-shell.
 Notice that such contribution is
not related to any gluonic exchange, differently from the usual perturbative
expansion
based on factorization theory \cite{CL} where the exchange of one gluon is
supposed to
be dominant. The main argument of Ref. \cite{CRS} is that it is possible to
write down a sum rule for Compton scattering in terms of a perturbative
spectral function
calculated (to lowest order) from Fig.~5a and from its power corrections.
On the other hand, the resonance region is modeled, as usual, in terms
of a single resonance (the double pole at $m_\pi^2$, the pion mass squared)
and a continuum contribution starting at a virtuality $s_0$.
Information on the invariant amplitudes of
pion Compton scattering can be obtained from the residue at the pion double
pole
of the correlator in (1).

In fact, the invariant amplitudes of pion Compton scattering
can be isolated from the matrix
element \beq
M_{\nu\lambda}= i\int d^4y \, e^{-iq_1 y}
  \langle p_2|T\left(J_{\nu}(y)J_{\lambda}(0)\right) |p_1 \rangle .
  \label{mnula}
\eeq

In Ref. \cite{CRS} we have expanded (3) in terms of $H_1$ and $H_2$
defined by
the relation
\beq
M^{\lambda\mu}= H_1(s,t) e^{(1)\lambda}e^{(1)\mu} +
H_2(s,t) e^{(2)\lambda}e^{(2)\mu},
\label{h1h2}
\eeq
where $e^{(1)}$ and $e^{(2)}$ are polarization vectors, which are defined
below.
The residue at the pion pole is isolated from the exact
spectral density  of (1) as
\beqa
\Delta_{\mu\nu\lambda\sigma}
&=&
f_\pi^2\, p_{1\mu}p_{2\sigma}\, (2\pi)^2\delta(p_1^2-m_\pi^2)
\delta(p_2^2-m_\pi^2)\nonumber \\
&\ & \times M_{\nu\lambda}(p_1,p_2,q_1)+\Delta^{cont} \,\,\,\,\,  + h.\,\, r.
\eeqa
where $M_{\nu\lambda}$ is given by eq. (4), $\Delta^{cont}$
is the double discontinuity of the multiparticle states
of the correlator in eq. (1), and $h.\,r.$
denotes the contribution due to the higher resonances.
In the derivation of the sum rule it is assumed that other single particle
states,
corresponding to the double poles located at momenta $p_1^2,p_2^2$ equal to the
masses of
higher resonances, give negligible contributions.
Therefore $\Delta_{\mu\nu\lambda\sigma}$
contains the most valuable non-perturbative information on the invariant
amplitudes $H_1$ and $H_2$
at the pion pole, although mixed with the contribution of the higher states.
While the contributions from $\Delta^{cont}$ - for virtualities bigger than
a threshold value $s_0$ - cancel from both sides of the sum rule (see eqs. (10)
and
(11) below),
a suitable way to remove the contribution of single particle states from (5),
still keeping the pionic contribution ($H_1$ and $H_2$), is to use a
projection.
 In Ref. \cite{CRS} this has been obtained using two projectors, defined in
terms
 of the vectors $e^{(1)}_\lambda ,e^{(2)}_{\lambda}$ and $n_{\lambda}$, the
last one
 satisfying the requirements
 \beq
n\cdot (q_1)=n\cdot(p_1-p_2)=n^2=0,
\label{ndef}
\eeq
 and possibly choosen as
\beq
n^\mu = \biggl (e^{(2)} \pm ie^{(1)} \biggr )^\mu\ .
\label{nconstruct}
\eeq
Improvement of the sum rule is obtained, as usual, by acting on it with
double Borel transforms \cite{SVZ,NR1,JS}.
In the case of pion Compton scattering
the two projected spectral densities are given by the
expressions \beqa
\Delta_4^{(12)}(p_i^2,s,t)
&=&
f_\pi^2\, n\cdot p_{1} n\cdot p_{2}\, (2\pi)^2\delta(p_1^2-m_\pi^2)
\delta(p_2^2-m_\pi^2) \nonumber \\
&\, & \times \left ( H_1(s,t)+
 H_1(s,t) \right )\, +  \Delta_4^{(12)cont},
 \eeqa
and
\beqa
\Delta_4^{(1)}(p_i^2,s,t)
&=&
f_\pi^2\, n\cdot p_{1} n\cdot p_{2}\, (2\pi)^2\delta(p_1^2-m_\pi^2)
\delta(p_2^2-m_\pi^2) \nonumber \\
&\, & \times  H_1(s,t)\, +  \Delta_4^{(1)cont}.
 \eeqa
In the following we shall assume $m_\pi$ to be zero.
Then the full sum rule or "master equation" \cite{CRS} for the invariant
amplitudes
of pion Compton scattering, for instance for
$H_1 + H_2$,
 can be written down as
\beqa
&&f_\pi^2\, n\cdot p_{1} n\cdot p_{2}
\left ( H_1(s,t)+
 H_2(s,t) \right ) \nonumber \\
&&\quad  = \int_{0}^{s_o}ds_1\int_{0}^{s_o}ds_2\ \Delta^{(12)pert}_{4}
(s_1,s_2,s,t)e^{-s_1/M_1^2 -s_2/M_2^2} \nonumber \\
&& \quad \quad
\left (1 - e^{-(\lambda^2-s_1)/M_1^2} \right ) \left (1 -
e^{-(\lambda^2-s_2)/M_2^2} \right)
+ \,\,\,\, power \,\,corr.\,,
 \label{compsr}
\eeqa
where $\lambda^2\equiv (s+t)/4$ is the radius of the contour in the spectral
representation
of the correlator (1)  (see Fig.~4)
 and the remaining terms in (10) are the power
corrections. In the local duality limit of eq.~(10) the power corrections are
neglected. Keeping $M_i^2$ and $\lambda/M_i^2$ to be large, the sum rule for
$H_1 + H_2$
takes the approximate form \beq
{f_\pi}^2 n\cdot p_1 n\cdot p_2 \left( H_1(s,t) + H_2(s,t))\right)=
\int_{0}^{s_0}ds_1 \int_{0}^{s_0}ds_2 \,\Delta^{(12)pert}(s_1,s_2,s,t),
\eeq
in which the exponentially suppressed terms present in eq. (10) have been
removed.
A similar sum rule can be written down in terms of only $H_1$ and
$\Delta^{(1)pert}$.
The asymptotic expansion at fixed angle of the sum rule for the
two perturbative
spectral densities $\Delta^{(1)pert}$ and $\Delta^{(12)pert}$, calculated from
the
ope-side of the sum rule,
has been given in \cite{CRS}. In the next section
we are going to discuss their complete evaluation to lowest order in
$\alpha_s$.

\section{Perturbative spectral functions}
In this section we present the complete evaluation of the lowest order
perturbative
spectral functions which appears in the sum rule for the sum
of the two helicity amplitudes eq.~(\ref{compsr})
and for the single invariant amplitude $H_1(s_1,s_2,s,t)$.
In a more complete analysis of the sum rule for pion Compton scattering,
an asymptotic expansion of the lowest order spectral functions is no longer
sufficient. In fact, while the dominant $O(1/Q^4)$ and the next to leading
contributions
can be obtained in a more simple way, by a simple power counting of the various
integrals,
the determination of the $O(1/Q^8)$ contributions requires a full scale
calculations
of all the terms from Fig. 1a.
This turns out to be a rather lenghty task.
The asymptotic expansion - arrested to $O(1/Q^4)$ - of the sum rule for $H_1 +
H_2$
has been already
given in Ref. \cite{CRS}. In that work it has been shown that, at fixed angle,
Compton scattering has a strong similarity to the form factor case.
This crucial observation has opened the road to the extension of
the sum rule programme to 4-point functions for Compton processes,
while providing a calculational scheme which is
a direct generalization of those known in the form factor case \cite{JS,NR1}.

The leading spectral functions $\Delta^{(1)pert}$ and $\Delta^{(12)pert}$
can be calculated in terms of 3-cut integrals of the form (see Fig. 5a)
\beq
\label{if}
I[f(k^2,k\cdot p_1,...)]=\int
d^4k \, f(k^2,k\cdot
p_1,...){\delta_+(k^2)\delta_+((p_1-k)^2)
\delta_+((p_2-k)^2)\over (p_1-k + q_1)^2},
\eeq
\\
\beq
\label{ipf}
I'[f(k^2,k\cdot
p_1,...)]=\int d^4k
f(k^2,k\cdot p_1,...)\delta_+(k^2)\delta_+((p_1-k)^2)\delta_+((p_2-k)^2), \eeq
%The contribution to $\Delta^{(12)pert}(s_1,s_2,s,t)$
%given by the expression \cite{CRS}
The analogy to the form factor approach (see Fig.~5b) is
clearly due to the fact that (\ref{if}) approximates
the total discontinuity (to lowest order in $\alpha_s$)
along the positive $s_1,s_2$ virtualities only
by a "3-particle cut", while one propagator (the top line of Fig.~5a) is kept
off shell.
The dominance of this discontinuity compared to other contributions can be
easily proved.
Other cuts are 1) either subleading or 2) they are relevant for forward
scattering (not considered in this work) or 3) they are not in the physical
region.
%\beqa
%\Delta^{(12)pert}&=&16 \,(s_1 -s + p_1 q_1 + p_2 q_1)\left( I[n\cdot k n\cdot
%%p_1] -
%I[(n\cdot k)^2] \right)  \nonumber \\
%&\ & \qwad \qwad -(p_1\cdot q_1 + p_2 \cdot q_1)
%\left( I'[(k\cdto n)^2] - I' [(k\cdot n)^2] \right). \\
%\eeqa
The evaluation of these integrals is not obvious and can be handled
in a suitable reference frame. It simplifies
considerably in the pion "brick-wall" frame
in which the momenta of the two pions
 are parameterized as
\beq
p_1=Q \bar{v} +{s_1\over 2 Q} v, \,\,\,\,\,\,\,\,\,\,\, p_2={s_2\over 2
Q} \bar{v} + Q v,
\eeq
where $v$ and $\bar{v}$ are light-cone momenta, such that
$v^2 = \bar{v}^2=0, \,\,\, v\cdot\bar{v}=0. $
The momentum transfer  $t$ can be expressed
in terms of $Q^2$, which acts as a large "parameter" in the scattering process,
by the relation

\beq
t=(p_2-p_1)^2=s_1+s_2 -2 Q^2- {s_1 s_2 \over 2 Q^2}.
\eeq
  $Q^2$ is given by
\beq
\label{qquadro}
Q^2={1\over 4} (s_1+s_2-t + \delta) = {1 \over 4} (s+u+\delta)\ ,
\eeq
with
\beq
\delta=\sqrt{(s_1+s_2-t)^2 -4 s_1 s_2} = {4 Q^4 - s_1 s_2\over 2 Q^2}.
\label{deltadef}
\eeq
In the frame specified above, all the transversal momentum is carried by the
two photons.
Notice that $t=-2 Q^2$ for $s_1=s_2=0$.
In our approach the two photons are on shell $(q_1^2,q_2^2=0)$
and carry physical polarizations. It should be noticed that the matrix
element in (3) admits various possible extensions
(in terms of polarization vectors) when we move off shell the two pions.
This extension must be done in such a way that the requirements of physical
polarizations for the two photons are still satisfied. A suitable choice
of the polarization vectors drastically simplifies the calculations.
The two polarization vectors that we are going to use in our calculations are
 slightly different from those given in \cite{CRS}. In our case they are
defined as
\beq
e^{(1)\lambda}={N^{\lambda}\over \sqrt{-N^2}} \,\,\,\,\,\,\,\,
e^{(2)\lambda}={P^{\lambda}\over \sqrt{-P^2}},
\label{edef}
\eeq

\beq
N^{\lambda}=\epsilon^{\lambda\mu\nu\rho}P_{\mu}r_{\nu}R_{\rho}
\,\,\,\,\,\,\,\,
P^{\lambda}=\nu_1 p_1^\lambda+ \nu_2 p_2^\lambda +{\alpha\over2} R^\lambda ,
\eeq
where
\beq
\nu_1 = {p_1\cdot p_2 - s_2} \,\,\,\,\,\,\,\nu_2=p_1\cdot p_2-s_1,
\eeq
and
\beq
\alpha=(2/t)(\nu_1 p_1\cdot R + \nu_2 p_2\cdot R),
\eeq

\beq
R=q_1+q_2 \ \ \ \ \ \ \ \ r=q_2 -q_1.
\eeq
In particular, the off-shell extension of $P^\lambda$ has been chosen
in eq. (19) to be
more symmetric compared to the case considered in \cite{CRS}, and includes two
coefficients, $\nu_1$ and $\nu_2$, which are equal to unity at the pion pole.
 Even with this new extension, the polarization vectors $e^{(i)}$
still satisfty the requirements
\beqa
e^{(i)} \cdot q_1 &=& e^{(i)}\cdot q_2 = 0, \nonumber \\
e^{(i)}\cdot e^{(j)} &=& -\delta_{ij}\
\label{eortho}
\eeqa
for physical polarizations.
These relations hold for all positive $s_1$ and $s_2$, whether
or not they are equal, and for $\nu_1$ and $\nu_2$ chosen as specified in eq.
(20).

The projector $n^\mu=(n^+,n^-,n_\perp)$, using the fact that in the "brick
wall" frame
\beq
e^{(1)}=(0,0,e^{(1)}_{\perp})\,\,\,\,\,\,\, \,\,\,\, e^{(1)}_{\perp} \cdot
q_\perp =0,
\eeq
takes the form
\beqa
n^\pm &=& \left(\nu_1 {p_1}^\pm + {\nu_2}{p_2}^\pm +{ \alpha\over 2}({q_1}^\pm
+
{q_2}^\pm) \right)/\sqrt{-P^2}, \nonumber \\
n_\perp &=&{\alpha q_\perp\over \sqrt{-P^2}} + i e_1.
\label{enne}
\eeqa
Notice that only the transversal part of $n^{\mu}$ is imaginary.
The advantage of using such a frame is due to the fact that the spectral
densities
 can be expressed as rational functions of the variables
$Q^2,s,s_1,s_2$. They are also polynomial in $Q^2$ and
 symmetric in $s_1$ and $s_2$, as expected
from the time reversal invariance of the scattering process.
This provides a drastic simplification of the calculations.
It is not hard to show that the same simplifications occurs
in the case of the pion form factor, once it is treated in the same reference
frame.

The final expression for $\Delta^{(12)pert} $ is given by
\beqa
\Delta^{(12)pert} &=&{5\over 3 (2 \pi)^3}
 \{ {8 Q^2 (s- 2 Q^2)R_{12}(Q^2,s,s_1,s_2)\over  (4
Q^4-s_1 s_2)^5  (2 Q^2 s - s_1 s_2)}  \nonumber \\
  &\ & \quad \quad \quad \quad \quad \quad +
  {8 Q^2 (u- 2 Q^2)R_{12}(Q^2,u,s_1,s_2)\over (4 Q^4 -s_1 s_2)^5(2 Q^2 u - s_1
s_2)}
\biggr \} ,\nonumber \\
\label{depert12}
\eeqa
where
\beq
R_{12}(Q^2,s,s_1,s_2) = \sum_{n=0}^8 b_n(s,s_1,s_2) Q^{2n}
\label{R12}
\eeq
is polynomial in $Q^2$, and with a similar expression for
$R_{12}(Q^2,u,s_1,s_2)$ obtained
replacing $s$ with $u$ in eq.~(\ref{R12}).
This second term comes from the crossed diagram of Fig.~1a, where the two
photon lines
are interchanged.

The coefficients  $b_n(s,s_1,s_2)$ and $b_n(Q^2,u,s_1,s_2)$ are
symmetric in $s_1$ and $s_2$.
Their expressions are given by
\beqa
b_0 &=& -s_1^5 s_2^5,  \nonumber \\
b_1 &=& 8 s_1^4 s_2^4 (s_1 + s_2),  \nonumber \\
b_2  &=& -4 s_1^3 s_2^3(5 s s_1 + 5 s s_2 +14 s_1 s_2),  \nonumber \\
b_3 &=& 16 s_1^2 s_2^2 (s^2 s_1 + s^2 s_2 + 14 s s_1 s_2 + 2 s_1^2 s_2 +
2 s_1 s_2^2),  \nonumber \\
b_4 &=& -16 s_1^2 s_2^2 (12 s^2 + 15 s s_1 + 15 s s_2 +4 s_1 s_2), \nonumber \\
b_5 &=& 128 s_1s_2 (2 s^2 s_1 + 2 s^2 s_2 +4 s s_1 s_2 + s_1^2 s_2 +s_1 s_2^2),
\nonumber  \\
b_6 &=& -192 s_1 s_2 (4 s^2 + s s_1 + s s_2 +2 s_1 s_2 )\nonumber \\
b_7 &=& 256 s (s s_1 + s s_2 +2 s_1 s_2),   \nonumber \\
b_8 &=& 256 (-s s_1 - s s_2 + s_1 s_2).\nonumber \\
\label{bes}
\eeqa
Using the fact that at the double pion pole
\beq
\label{pole}
(n\cdot p_1)( n\cdot p_2) = s (s+t)/(-t),
\label{np}
\eeq
and inserting this result into eq. (11), we can write down
a local duality sum rule for $H_1 + H_2$ as
\beq
f_\pi^2 \left ( H_1(s,t ) + H_2 (s,t)\right ) = \int_{0}^{s_0}ds_1
\int_{0}^{s_0}ds_2
\,\rho^{(12)pert}(s_1,s_2,s,t),
\eeq
 with a spectral density given by
\beq
\rho^{(12)pert}={(-t)\over s
(s+t)}\Delta^{(12)pert}(s_1,s_2,s,t). \eeq

By a simple power counting in $Q^2$, one can immediately notice that the
asymptotic behaviour of the
spectral functions, as given by (\ref{depert12})  show a suppression by one
additional power of t ($t\approx -2 Q^2$,  specifically $O(1/t^2 )$) compared
to the
perturbative result predicted by the dimensional counting rules \cite{BL}.
This suppressed
power-law behaviour, as discussed in \cite{CRS}, is similar to the one obtained
for the pion form factor
and is not related to any gluonic exchange since the sum rule starts at
$O(\alpha_s^0)$.
The contributions of the power corrections
 (Figs.~2,3), as we are going to discuss next, start at
$O(1/Q^6)$ and can be arrested at $O(1/Q^8)$.
It turns out that a calculation of the power corrections, up to this order,
can be done without a full evaluation of such diagrams (see appendix  B).

\section{Modified Borel transform and gluonic corrections}

	One additional complicacy which appear in the case of 4-point functions,
 compared to 2 and 3 point functions, is related to the fact that the usual
definition
 of the Borel transform, so largely employed in the sum rule context,
 is no more valid. The reason is quite elementary and is essentially due to the
fact
 that only a finite region of analiticity is allowed for the correlation
functions.
 At fixed angle, such region can be depicted as a circle (see Fig. 4)
 and application of the ope is realistically possible
 only if the bigger radius of the contour
 is large enough \cite{CRS}. The ope then descibes the behaviour of the
correlation function
 exactly on the contour (the far left side of it, also termed "the deep
euclidean
 region"). Here we are going to assume that only the first lowest dimensional
operators
 are relevant for our needs.
Assuming a validity of the ope, implemented with only the lowest
dimensional condensates, we can proceed to the evaluation of such
non-perturbative
corrections.
The coefficients (Feynman integrals)
of such lowest dimensional operators are, however, analytic
functions of the pion virtualities inside the circle in Fig.~4
 and can be reexpressed in the form of contour integrals
in the $s_1,s_2$ complex planes.
Next we are going to show that such integrals can be approximated by their
double discontinuities parts along the cut
from zero to $\lambda^2$ in the $s_1$ and $s_2$ planes,
integrated on a square of size
$\lambda^2$.
To be specific, let's consider in fact the integral operator
\beqa
\tilde{B}(p_1^2\to M_1^2,p_2^2\to M_2^2)&=&{1\over (2\pi i)^2}\int_C
{dp_1^2\over
M_1^2}\int_C {dp_2^2\over M_2^2} \, e^{-p_1^2/M_1^2 - p_2^2/M_2^2}\nonumber \\
& & \quad \times \left(1-e^{(\lambda^2-s_1)/M_1^2}\right)\left(
1-e^{-(\lambda^2-s_2)/M_2^2}\right),\nonumber \\
&&
\label{newborel}
\eeqa
where $C$ denotes any contour that encloses the branch cut in the $p_1^2$ and
$p_2^2$
planes (the shadowed region in Fig.~4), from zero to $\lambda^2$, the radius of
the region
of analiticity for the correlator of 4-currents, and let's act with it on any
scalar
integral of the form $T_(p_1^2,p_2^2,s,t)$, defined by a Cauchy representation
of the form
\beq
T_(p_1^2,p_2^2,s,t)=-{1\over 4 \pi^2}\int_{\gamma_1}ds_1\int_{\gamma_2}
{T_(s_1,s_2,s,t)\over {(s_1-p_2^2)(s_2-p_2^2)}}
\label{t4g},
\eeq
with a possible choice of $\gamma$ depicted in Fig.~4.
If we let $\Delta_4(s_1,s_2,s,t)$ denote its double discontinuity across the
cut,
then, using eq.~(\ref{newborel}),  we easily obtain
\beqa
T_4(M_1^2,M_2^2,s,t)&=&\tilde{B}(p_1^2 \to M_1^2,p_2^2\to M_2^2)
T_4(p_1^2,p_2^2,s,t)
\nonumber \\
&=& \int_{0}^{\lambda^2} ds_1\int_{0}^{\lambda^2}ds_2
\Delta(s_1,s_2,s,t)e^{-s_1/M_1^2
-s_2/M_2^2} \nonumber \\
& & \quad \quad \quad \times \left( 1- e^{-(\lambda^2 - s_1)/M_1^2}\right)
\left( 1- e^{-(\lambda^2-s_2)/M_2^2} \right).\
\label{bt}
\eeqa
Eq.~(\ref{bt}) shows that the new transform, acting on each coefficient of the
power
corrections,  gives as a result a double integral along the virtuality cut from
zero to
$\lambda^2 $, times an additional exponential suppression  at the edge of the
integration
region.  Notice that such additional factors have been introduced  into the
definition of the
transform  eq.~(\ref{newborel}) in order to ensure that the crossing between
the two contours
$C$ and $\gamma$ in Fig.~4  in eqs.~(\ref{t4g}) and (\ref{newborel})
 does not induce any singularity for $s_1,s_2=\lambda^2$.
We also notice that  the  influence of such  additional suppression on the
explicit expression
of the sum rule is likely to be modest, since the  usual values of the Borel
mass,
obtained from the stability analysis of the sum rule for the pion form factor,
 are of the order of  1 $Gev^2$, while $\lambda^2$ is  supposed to be higher.
Therefore, the
new definition of the transform compares favourably well with  the usual one
given by \cite{SVZ}

\beq
B(Q^2\to M_1^2)=lim_{\stackrel{Q^2,n\to \infty}{Q^2/n=M_1^2}}\frac{1}{(n-1)!}
(Q^2)^n (-\frac{d}{dQ^2})^n.
\label{dif}
\eeq
In the case $\lambda^2\to\infty$, $\tilde{B}(Q^2\to M_1^2)$ and (\ref{dif})
coincide.
  From eq.~(\ref{bt}) we come to the conclusion that, from each
Feynman integral which appears in the evaluation of the power corrections, the
modified Borel
transform picks-up only the part of it integrated along the cut, and therefore
we are allowed
to neglect all  the complex contributions in eq.~(\ref{t4g}).
  The method we implement here
 is  a generalization of the one adopted in the simpler case of three-point
functions
\cite{JS}. The next point to address in the calculation of the power
corrections to the
diagram in Fig. ~1 is which insertion of condensates should be included.  Our
arguments
on this last point are entirely based on the similarity between the pion form
factor  and
Compton scattering.  As above, in our kinematic condition, characterized by
 $s,t,u$ of moderate size, all of them in the physical region, and at
fixed $s/t$, operator insertions on the quark line connecting the two photons
in Fig.~1 are in
clear disagreement with this similarity picture.
By the same token, all the operatorial insertions which lower the virtuality of
the quark line
connecting the two photon lines in the diagrams of the power corrections will
not
be kept into account.  A typical example is given in Fig.~2b, where the cutting
of four quark
lines and the specific attachment of an internal gluons guarantees continuity
of the momentum
flow in the loop,  but lowers the virtuality of the quark line connecting the
two
photons at the top.
Now we come to discuss the practical implementation of the method.
In the case of the power corrections, the general tensor operator
$G_{\mu\nu}G_{\alpha\beta}$
has been projected  onto its corresponding scalar average ($\langle
G^2\rangle=\langle
G^{a}_{\mu\nu}G^{a \mu\nu}$) by the
formula \beq
\langle 0|G_{\mu\nu}G_{\alpha\beta}|0\rangle ={1_c\over 72 }\langle 0|
G^{a}_{\rho\sigma}G^{a \rho\sigma}|0\rangle (g_{\mu\alpha}g_{\nu\beta}
-g_{\mu\beta}g_{\nu\alpha}).
\label{ave}
\eeq
where $1_c$ is the unit $3\times 3$ color matrix.
We also assume that
\beq
G_{\mu\nu}=g_s G^{a}_{\mu\nu}{\lambda^a\over 2}\,\,\,\,\,\, \,\,\,\,
\alpha_s={g_s^2\over 4 \pi}
\label{rage}
\eeq
with $\lambda^a$ $(a=1,2,...8)$ denoting the standard Gell-Mann matrices.
Eq.~(\ref{ave}) can be also rewritten in the form
\beq
\langle 0|G^a_{\mu\nu}G^b_{\alpha\beta}|0\rangle ={1\over 96 }\langle 0|
G^{a}_{\rho\sigma}G^{a \rho\sigma}|0\rangle (g_{\mu\alpha}g_{\nu\beta}
-g_{\mu\beta}g_{\nu\alpha})
\eeq
by using standard properties of the Gell-Mann matrices.

As usual we work in the Fock-Schwinger gauge
 defined by the condition
\beq
z^\mu A_\mu(z)=0
\eeq
and we employ a simplified
momentum-space method of evaluation of the Wilson coefficients discussed in
Ref. \cite{mallik}.
In the case of massive quarks, the fermion propagator in the external
gluonic background is given by \cite{mallik}
\beqa
S(p) &=& {\not{p} +m\over p^2 - m^2}\,\, + \,\,{1\over 2} i
{(\gamma^{\alpha}\not{p}
\gamma^{\beta} G_{\alpha \beta}
 - m \gamma_{\alpha}G^{\alpha\beta}\gamma_{\beta})\over (p^2
- m^2)^2} \nonumber \\
&\ &  \quad \quad + {\pi^2 \langle G^2\rangle m\not{p} (m +\not{p})\over (p^2 -
m^2)^4},
\label{propagator}
\eeqa
which simplifies considerably in the massless limit.
Notice that in this limit the third term of eq.~(\ref{propagator}),
corresponding to the
insertion of two external gluon lines
vanishes and only diagrams with a single gluonic insertion on each fermion line
(the second term of eq.~(\ref{propagator})) survive.

In order to better illustrate our approach (see also the discussion contained
in Appendix B), let's start considering
 the scalar integral
\beqa
&& J(p_1^2,p_2^2,s,t,m_1,m_2) \nonumber \\
&& = \int {d^4k  \over
(p_1-k)^2 (k^2-m_1^2)^2 \left((p_2 -k)^2 - m_2^2\right)^2 (p_1 - k + q_1)^2}
\nonumber \\
&& ={\partial\over \partial m_1^2}{\partial\over \partial m_2^2}
\tilde{J}(p_1^2,p_2^2,s,t,m_1,m_2)
\label{factorm}
\eeqa

where we have defined
\beq
\tilde{J}(p_1^2,p_2^2,s,t,m_1,m_2)=\int
{d^4k\over (k^2 - m_1^2)\left( (p_2-k)^2 - m_2^2\right) (p_1-k)^2 (p_1-k +
q_1)^2}\ .
\label{jmo}
\eeq

The dispersive part of eq.~(\ref{jmo}) can be approximated as
\beq
\label{jm1}
J(p_1^2,p_2^2,s,t,m_1,m_2)=-{1\over 4
\pi^2}\int_{0}^{\lambda^2}ds_1\int_{0}^{\lambda^2}ds_2
\,{\Delta^{scalar}(s_1,s_2,s,t,m_1,m_2)\over (s_1 - p_1^2)(s_2 - p_2^2)}\
\eeq
with a spectral density given by
\beqa
&&\Delta^{scalar}(s_1,s_2,s,t,m_1,m_2) \nonumber \\
&& =(-2\pi i)^3\int d^4k\,
{\delta_+(k^2-m_1^2)\delta_+((p_2-k)^2 - m_2^2)\delta_+((p_1-k)^2)\over (p_1-k
+q_1^2)}
\label{disc}
\eeqa

 The dispersive part of the scalar integral associated to eq.~(\ref{factorm})
when we set $m_1$ and $m_2$ to be zero
can be written down in the following form
\beq
J(p_1^2,p_2^2,s,t)
\equiv \int {d^4 k \over (p_1-k)^2 \left( k^2\right)^2 \left( (p_2 -k)^2
\right)^2 (p_1 - k + q_1)^2} \nonumber
\eeq
\beq
\label{disc1}
=-{1\over 4 \pi^2}\int_{0}^{\lambda^2}ds_1 \int_{0}^{\lambda^2}ds_2\,{
\bar{\Delta}(s_1,s_2,s,t)\over (s_1 - p_1^2)(s_2-p_2^2)},
\eeq
where we have defined

\beq
\label{disc2}
\bar{\Delta}(s_1,s_2,s,t)\equiv \left({\partial\over \partial
m_1^2}{\partial\over
 \partial m_2^2}\Delta^{scalar}(s_1,s_2,s,t,m_1,m_2)\right)_{m_1,m_2=0}
\eeq

Evaluation of (\ref{disc}) gives
\beq
\label{delm}
\Delta^{scalar}(s_1,s_2,s,t,m_1,m_2)=(-2\pi i)^3 {\pi\over 2 \delta W(m_1,m_2)}
\eeq
with
\beqa
W(m_1,m_2) &=& - 4 m_1^2 (Q^2)^2 + 2 m_2^2 Q^2 s + 4 (Q^2)^2 s + 2 m_1^2 Q^2
s_1 -2 m_2^2 Q^2
s_1\nonumber \\
& & -2 Q^2 s s_1 + 2 m_1^2 Q^2 s_2 - 2 Q^2 s s_2 - m_1^2 s_1 s_2 + s s_1 s_2
\label{w}
\eeqa
and with $\delta$ given as in eq.~(\ref{deltadef}).
Inserting this result in eq.~(\ref{delm}) and using (\ref{disc2}) we get
\beqa
\bar{\Delta}(s_1,s_2,s,t)={-32 \pi^4i(Q^2)^2 (s_1 -s)(-4 (Q^2)^2 +2 Q^2 s_1 - s
s_1
+ 2 Q^2 s_2)\over s^3 (2 Q^2 - s_1)^3 (-2 Q^2 + s_2)^3}\ .
\label{disc3}
\eeqa
It is easy to show that the method works perfectly well for  values of $m_1$
and $m_2$
small compared to  $Q^2$.  In fact the spectral functions so obtained are
easily shown to be
regular  when the virtualities of the two external pions vary independently
 in the $(0,\lambda^2)$ interval. This fact insures are that no poles are
generated.
However, even though the calculation scheme we have developed
reduces to few steps, its practical implementation requires considerable
effort.

The 3 direct diagrams in Fig. 3 have been evaluated by this method.
An asymptotic expansion of each coefficient has been made.
We get
\beqa
\Delta^{(12)3a}_{gl}&=&{5\alpha_s\over 54 \pi^2}
\langle G^2\rangle {(s-2 Q^2)Q^4 (s-s_1)^2 (2 Q^2 s
- s_1 s_2)\over s^3 (2 Q^2 - s_1)^3 (2 Q^2 - s_2)^3}
\label{pc1}
\eeqa
\beqa
\Delta^{(12)3b}_{gl}&=&{20\alpha_s\over 27 \pi^2}\langle G^2\rangle {(2
Q^2-s))( Q^4
-4 Q^2 s + s^2 + s s_1 +4 Q^2 s_2)\over
s (2 Q^2 - s_1)^4(2 Q^2 - s_2)^4}  \label{pc2}
\eeqa
 \beqa
 \Delta^{(12)3c}_{gl}&=&{5\alpha_s\over 54\pi^2} \langle G^2\rangle {Q^6(-4 Q^4
+2 Q^2
s - Q^2 s_1 + s s_1 -3 Q^2 s_2 + s  s_2)\over
 s^3 (2 Q^2 -s_1)^3 (-2 Q^2 +s_2)^3} \nonumber \\
 &&
\label{pc3}
\eeqa
where the average on the external gluonic field has been made by using eq.
(\ref{ave}).

Crossed diagrams with interchanged photon lines have been treated with the same
method
discussed in this section. However, a much faster way to
obtain their contributions to the power corrections
is by using the $s\to u$ crossing
properties of the spectral function $\Delta^{(12)}$. In terms of the
variables $Q^2,s$ and
$t$ this is equivalent to the substitutions

\beqa
Q^2&\to& Q^2; \nonumber \\
s &\to& u= s_1 + s_2 -s - t ;\nonumber \\
t &=& s_1 + s_2 -2 Q^2 - {s_1 s_2\over 2 Q^2}.
\label{qs}
\eeqa
in eqs.~(\ref{pc1},\ref{pc2},\ref{pc3})
Combining the contributions from eqs.~(\ref{pc1},\ref{pc2},\ref{pc3}),
and adding the crossed contributions we can finally represent
the gluonic power corrections to the spectral
function into the form
\beq
\label{pcgluon}
\Delta^{(12)}_{gluon}={20\alpha_s\over 27 \pi^2}{
Q^8\tau^{gluon}(Q^2,s,s_1,s_2)\over
s (2 Q^2 -s_1)^4  (2 Q^2 -s_2)^4 }\
\eeq
where
\beqa
\tau^{gluon}(Q^2,s,s_1,s_2) &=&8 Q^6 -24 Q^4 s +14 Q^2 s^2 -2 s^3 +7  Q^4
s_1\nonumber \\
& &\quad   +2 Q^2 s s_1 -3 s^2 s_1 +17 Q^4 s_2 -6 Q^2 s s_2 -s^2 s_2 \nonumber
\\
 & &
 \label{gluonic}
\eeqa
This result shows that power corrections for the new sum rules derived in Ref.
\cite{CRS}
are calculable in a systematic form in full analogy to the form factor case.
The power-law fall-off of such contributions is $O(1/Q^6)$, two additional
power of $Q$
suppressed compared to the perturbative contribution.

\section{Quark power corrections and $\rho^{(12)}$.}
Quark power corrections can be calculated by the usual methods \cite{JS}.
Typical contributions of this type
are depicted in Figs.~2. Let's focus our attention on these last diagrams,
in which the 4 broken lines simbolize a soft insertion of a matrix element
proportional
to $\langle (\bar{\psi}\psi)^2\rangle $. In the limit of massless quarks the
contribution proportional to
 $\langle (\bar{\psi}\psi)\rangle $ are zero.
As usual \cite{JS}, the average of the 4-quark operators can be obtained from
the standard expression
\beq
\langle 0|\bar{\psi}^a_\alpha\bar{\psi}^b_{\beta}\psi^c_\gamma
\psi^d_\delta|0\rangle
={1\over 144}\langle (\bar{\psi}\psi)^2\rangle
(\delta^{ad}\delta^{bc}\delta_{\alpha\delta}\delta_{\beta\gamma}-
\delta^{ac}\delta^{bd}\delta_{\alpha\gamma}\delta_{\beta\delta}).
\eeq

Combining all possible insertions which satisfy the requirements of leaving the
top quark
line connecting the two external photons virtual and after adding all the
crossed terms we
get \beqa
\Delta^{(12)quark}(s_1,s_2,s,t)&=&
{10\pi \alpha_s\tau^{quark}(Q^2,s_1,s_2)\over
   27 Q^2 s s_1^3 (-2 Q^2 + s_1)^2 s_2^3 (-2 Q^2 + s_2)^2},\nonumber \\
&&
\label{dequo}
\eeqa
where
\beqa
\tau^{quark}&= &(-2 Q^2 + s) (2 Q^2 s - s_1 s_2) (s_1^2 +
s_2^2) (16 Q^8 - 8 Q^6 s - 8 Q^4 s^2 \nonumber \\
&&- 16 Q^6 s_1 + 12 Q^4s s_1 +    4 Q^4 s_1^2 - 16 Q^6 s_2 + 12 Q^4 s s_2
\nonumber \\
&&+ 16 Q^4 s_1 s_2 -
      2 Q^2 s s_1 s_2 - 4 Q^2 s_1^2 s_2 + 4 Q^4 s_2^2 - 4 Q^2 s_1 s_2^2 +
s_1^2
s_2^2)\nonumber \\
 &&
 \label{tquark}
\eeqa

Adding together  eqs.~(\ref{depert12},\ref{pcgluon},\ref{dequo})
the operator product expansion of such spectral function is finally written
down in the form
\beqa
&&\rho^{(12)}(s_1,s_2,s,t)\nonumber \\
&&= \left({-t\over s
(s+t)}\right)\Delta^{(12)}(s_1,s_2,s,t)\nonumber \\
&&= \left({-t\over s
(s+t)}\right)\{\Delta^{(12)pert}(s_1,s_2,s,t)\theta(s_0-s_1)\theta(s_0-s_2)\nonumber \\
&& + \alpha_s\langle G^2\rangle{20\over 27 \pi^2}{
Q^8\tau^{gluon}(Q^2,s,s_1,s_2)\over
s (2 Q^2 -s_1)^4  (2 Q^2 -s_2)^4 }\nonumber \\
 && +{2\over
9}\langle (\bar{\psi}\psi)^2\rangle  {10\pi
\alpha_s\tau^{quark}(Q^2,s_1,s_2)\over
   27 Q^2 s s_1^3 (-2 Q^2 + s_1)^2 s_2^3 (-2 Q^2 +
s_2)^2}\delta(s_1-p_1^2)\delta(s_2-p_2^2)\},\nonumber \\
&&
 \label{ro12}
\eeqa
where we have included a factor 1/($n\cdot p_1 \,n\cdot p_2$) in the
normalization of the
spectral density and  we have used eq.~(\ref{np})
 As we already mentioned in the previous section, our calculations provide us
with a spectral density $\rho^{(12)}(s_1,s_2,s,t)$ which includes an exact
evaluation
of the lowest order contribution (Fig.~1) and is asymptotic in $1/Q$,
for the power corrections.
Eq.~(\ref{ro12}), once it is integrated  along the  $(0,\lambda^2)$ cut in each
of the
two virtualities $s_1$ and $s_2$, gives a sum rule of the form
\beq
f_\pi^2\left({ H_1(s,t) +H_2(s,t)\over p_1^2
p_2^2}\right)=\int_{0}^{\lambda^2}ds_1\int_{0}^{\lambda^2}ds_2{\rho^{(12)}(s_1,s_2,s,t)\over
(s_1-p_1^2)(s_2-p_2^2)},
\label{withdeno}
\eeq
where we have included all the contributions to the sum rule under a single
integral sign.
In order to further simplify eq.~(\ref{ro12}) in a consistent way, as we
already discussed
in the previous section, only 3 out of
the 8 coefficients given in eq.~(\ref{bes}), namely $b_8,b_7$ and $b_6$ are
needed.
By doing this approximation in (\ref{ro12})  and  using  eq.~(\ref{depert12})
it is
possible to write down the complete asymptotic spectral function, which
includes the  lowest
dimensional power corrections as
\beqa
&& \rho^{(12)}(s_1,s_2,s,t)\nonumber \\
&&\left({-t\over s
(s+t)}\right)\{ {320 Q^{14}\tau^{pert}(Q^2,s_1,s_2)\over 3 s (s-2 Q^2 )(4 Q^4 -
s_1 s_2)^5 (-2
Q^2 s + s_1 s_2)}\theta(s_0-s_1)\theta(s_0-s_2)\nonumber \\
 &&\quad +\alpha_s \langle G^2\rangle {20\over 27 \pi^2}{
Q^8\tau^{gluon}(Q^2,s,s_1,s_2)\over
s (2 Q^2 -s_1)^4  (2 Q^2 -s_2)^4 }\nonumber \\
&&\quad +\langle (\bar{\psi}\psi)^2\rangle  {10\pi
\alpha_s\tau^{quark}(Q^2,s_1,s_2)\over
   27 Q^2 s s_1^3 (-2 Q^2 + s_1)^2 s_2^3 (-2 Q^2 +
s_2)^2}\delta(s_1-p_1^2)\delta(s_2-p_2^2)\},\nonumber\\
&&
\label{thesum}
\eeqa
where
 \beqa
\tau^{pert}&=& 4 Q^4 s s_1 - 4 Q^2 s^2 s_1 +4 Q^4 s s_2 -4 Q^2 s^2 s_2
\nonumber \\
&& \quad -4 Q^4 s_1 s_2 -8 Q^2 s s_1 s_2 + 12 s^2 s_1 s_2
\label{tpert}
\eeqa
is the polynomial which appears as a factor in the numerator of the asymptotic
expansion
of the
perturbative contribution (the first term in eq.~(\ref{ro12})).
Notice that we still have to act  on eq.~(\ref{withdeno})  with the Borel
transform
eq.~(\ref{newborel}) in order
to obtain the usual Borel sum rule.  In order to show the applicability of the
new transform
and  to characterize its action  on the power corrections to the spectral
function, we need
some further discussion.

\section{The new Borel sum rule and the pion form factor.}
In this last section we discuss two points,
namely the  possibility to act on the sum rule in eq.~(\ref{withdeno})
 with the modified
Borel transform eq.~(\ref{newborel}) in order to get an expression which  has a
similarity with the usual sum rule for the pion form factor \cite{NR1,JS}, at
the same
time we will argue that the non perturbative power corrections are likely to
grow
much weaker compared to the form factor case
in the new sum rule for pion Compton scattering. We start from the first issue.

It is easily observed that the spectral density obtained in eq. (\ref{thesum})
is regular all over the integration region of the two pionic virtualities $s_1$
and$s_2$.
This is a nice feature of our result,  mainly due to the kinematic constraints
coming
from the fact that we are  investigating the process at fixed angle and with
moderate size
$s,t$ and $u$ invariants. In fact $Q^2$, as given by eq.~(\ref{qquadro})
is much larger than any of the two virtualities even at the edge of the
integration region
(see Fig.~4) when they are approximately equal to $(s+t)/4$.
This fact has an immediate consequence on the evaluation of the (modified)
Borel transform
of the quark power corrections, since the only poles which are encountered in
the
shadowed region in Fig.~4 by the new integral transform are those located at
$p_1^2,p_2^2=0$.
The evaluation of the residue can be obtained, as usual,
 by expanding  the overall factor
which multiplies the multiple pole $1/(p_1^2 p_2^2)^3$ in the expression of
the quark power corrections eq.~(\ref{dequo}) up to quartic terms in the
virtualities
of the two pions. This can be systematically done using eqs.~(\ref{qquadro},
\ref{deltadef})
to reexpress all the powers of $Q^2$ in terms of $s_1,s_2$ and $s$ and $t$.
By doing so and by
acting afterwards  with the modified Borel transform (\ref{newborel})  (with
$M_1=M_2=M$) on the resulting expression we get
\beqa
&&\tilde{B}(p_1^2\to M^2,p_2^2\to
M^2)\Delta^{(12)quark}(p_1^2,p_2^2,s,t)\nonumber \\ &&=\pi
\alpha_s\left(F_0(M,s,t) +
F_1(M,s,t)e^{-\lambda^2/M^2} + F_2(M,s,t)e^{-2\lambda^2/M^2}\right),\nonumber
\\
&&
 \label{quarkborel}
\eeqa
where
%\newpage
\beqa
&&F_0(M,s,t)\nonumber \\
&&={80\over 27}
{(-4 M^4 s - 8 M^2 s^2 -2 s^3 -2 M^2 s t - s^2 t + 2 M^2 t^2 +2 s
t^2 + t^3)\over M^4 t^2}, \nonumber \\
&&F_1(M,s,t)= {8\over 27} {(2 M^2 + s +t)(4 M^2 s + 2 s^2 - s t - t^2)
\over M^4 t^2}\nonumber \\
&& \\
&&F_2(M,s,t)= -{320\over 27}{s\over t^2}.
\eeqa
After that this intermediate step is done, we are finally in condition to write
down a
new sum rule for the sum of the two invariant amplitudes  $H_1$ and $H_2$ in
the
 helicity base as
\beqa
&&{f_\pi}^2\left(H_1(s,t) +H_2(s,t)\right)=\left({-t\over s(s+t)}\right)
\nonumber \\
&&\times \{\int_{0}^{s_0}ds_1\int_{0}^{s_0}ds_2 {320 Q^{14}\tau^{pert}
(Q^2,s_1,s_2)\over 3 s (s-2 Q^2 )(4 Q^4 - s_1 s_2)^5 (-2
Q^2 s + s_1 s_2)}\nonumber \\
 &&\quad + {20\over 27 \pi^2}\alpha_s\langle
G^2\rangle\int_{0}^{\lambda^2}ds_1\int_{0}^{\lambda^2}ds_2
 { Q^8\tau^{gluon}(Q^2,s,s_1,s_2)\over
s (2 Q^2 -s_1)^4  (2 Q^2 -s_2)^4 }e^{-(s_1+s_2)/M^2}\nonumber \\
&&\times \left(1- e^{-(\lambda^2-s_1)/M^2}\right)
\left(1- e^{-(\lambda^2-s_1)/M^2}\right)\nonumber \\
&&\quad +\pi\alpha_s\langle (\bar{\psi}\psi)^2\rangle\left( F_0(M,s,t) +
F_1(M,s,t)e^{-\lambda^2/M^2}
 + F_2(M,s,t)e^{-2\lambda^2/M^2} \right) \} \nonumber\\
&&
\label{suma}
\eeqa
where the expressions of  the quark and gluon spectral densities are given in
eqs.~(\ref{thesum}).
Our results has to be compared with its closest one, the sum rule for the pion
form factor.
In this case the usual gluon and quark power corrections give integrals which
can be explicitely performed
after that the standard Borel transform has been taken. One gets \cite{JS,NR1}
\beqa
&& f_\pi^2 F_\pi(Q^2)=\int_{0}^{s_0}ds_1\int_{0}^{s_0}ds_2\rho_{\pi
3}(s_1,s_2,t)e^{-(s_1+s_2)/M^2}
\nonumber \\
&& +{\alpha_s \langle G^2\rangle\over 12\pi M^4} +
\langle\left(\bar{\psi}\psi\right)^2\rangle {208\over 81}\pi\alpha_s\{1 +
{2\over 13}{Q^2\over M^2}\} +\,\,\, \,\, O({1\over M^8}) \,\, + O(\alpha_s)
\eeqa
where $\rho_{\pi 3}$ has been given in \cite{NR1} in the form
\beqa
&&{\rho_\pi}^{pert}_3(s_1,s_2,t) \nonumber \\
&& \quad \quad =
{3\over 2 \pi^4} t^2 \biggl( \left ({d\over dt}\right )^2 +
{t\over 3} \left ({d\over dt}\right )^3 \biggr)
{1\over ((s_1 + s_2 - t)^2 - 4 s_1 s_2)^{1/2}}
\label{ropi}
\eeqa

As we can see from this last equation, the power corrections tend to grow
rather
quickly in the form factor case and invalidate the sum rule at higher values
(beyond
$4,\, 5 \,\,Gev^2$), while the perturbative contributions have a rather fast
decay
$(1/Q^4)$.  This behaviour of the sum rule sets tight constraints on the range
of its validity.
The behaviour of the perturbative term in (\ref{suma}) is similar to the form
factor case, but the power correction are much more stabilized. This fact is
related to
the new angular dependence of the sum rule and to the suppression generated at
the
photon vertex (Fig. 5a). However,  in order to give a more precise and
definitive answer
to this last issue, a stability analysis  of the new sum rule and a
determination of the
Borel mass $M$ are required.

  \section{Conclusions}

We have extended
previous work on  local duality sum rule for pion Compton scattering  and
we have given  a complete derivation of the leading spectral function
 for a specific combination of the two invariant amplitudes in
the helicity base, $H_1 + H_2$,  showing that pion
Compton scattering can be systematically treated as in the pion form factor
case.
The final results for such function
turns out to be much more complicated compared to the case of the pion
form factor, but  they still have, at moderate momentum transfer,
 the same power law fall-off,
We have seen that the sum rule can be organized in the Breit frame of the
pion in a rather compact, albeit covariant, form.
We have also shown that the leading non perturbative corrections to the sum
rule are
calculable in a systematic way.  For this purpose we have found convenient to
use some
properties of a modified Borel transform which acts very simply on all the
terms of the sum rule and picks up the residue at the pole
of the pion virtualities
in the quark power corrections.
 A full discussion of the sum rule for the helicity amplitude $H_1$
and the stability analysis of the results contained in this work will be
considered in a separate work \cite{nuovo}.
Finally, we expect that our results and methods will find applications
to a new entire class of elastic hadron-photon interactions at intermediate
energy
and momentum transfer.
\vbox{\vskip 0.2 true in}
\centerline{Acknowledgements}
 I am most grateful to Prof. G. Sterman for suggesting this investigation,
for encouragement and for
discussions and to Profs. A. Radyushkin  for discussions.
I  acknowledge fruitful conversations with Profs. T. H. Hansson and H.
Rubinstein.
 I thank  Prof. G. Marchesini for discussions
and for  hospitality at Parma University, where part of this work has been
done; to   H. N. Li and to  R. Parwani for
correspondence  and to the Nuclear and Particle
Theory Group at CEN-Saclay for hospitality during a recent visit.

The large scale calculations involved in this work have been performed in part
using STENSOR by Lars H\"{o}rnfeldt, MATHEMATICA  by Wolfram Inc.,
TRACER by M. Jamin and M.E. Lautenbacher and by various packages
written by the author. I thank Lars H\"{o}rnfeld for assistance in using his
program.

This work was supported in part by the U.S. National Science Foundation,
grant PHY-9211367.

%\newpage
\renewcommand{\theequation}{A.\arabic{equation}}
\setcounter{equation}{0}
\vskip 1cm \noindent
\noindent {\large\bf Appendix A. Kinematics. }
\vskip 3mm \noindent

%\section{Appendix.Kinematics}
Let $q_1$ and $q_2$ be the momenta of the incoming and outgoing photon,
respectively, which are assumed to be on-shell ($q_1^2=q_2^2=0$).
Let also $p_1$ and $p_2$ be the momenta of the incoming and outgoing pion.
The external pion states are off-shell and are characterized by the
invariants $s_1=p_1^2$ and $s_2=p_2^2$.

We define as usual the Mandelstam invariants
\beq
s=(p_1+q_1)^2 \,\,\,\,\,\,\,\,\,\, t=(q_2-q_1)^2
\,\,\,\,\,\,\,\,\,,\,\,
u=(p_2-q_1)^2 \eeq
and the usual relation

\beq
s+t+u=s_1+s_2.
\eeq
%We consider both $s$ and $t$ to be very large and in the physical region
%($s>0,t<0$ ).

 It is also convenient to introduce
light cone variables for the photon momenta as follows
\beqa
q_1 &=& q_1^+ \bar{v} +q_1^- v +q_{1\perp},   \nonumber\\
 \bar{v}&=&{1\over\sqrt{2}} (1,1,{\bf 0}_\perp), \quad \quad \quad
 v={1\over\sqrt{2}} (1,-1,{\bf 0}_\perp) \nonumber \\
 q_{1\perp}\cdot n^{\pm} &=& 0.
\eeqa
In our conventions
\beq
v^+= {1\over \sqrt{2}} (v^0 + v^3) \,\,\,\,\, \bar{v}={1\over \sqrt{2} }(v^0 -
v^3).
\eeq

In the Breit frame of the incoming pion we also easily get
\beq
u=(p_2-q_1)^2=2 Q^2-s +{{s_1 s_2}\over 2 Q^2}.
\eeq

Covariant expressions for $q_1^\pm$ and $q_2^\pm$ can also be found
in the form
\beq
q_1^+={(s-2 Q^2)(2 Q^2-s_2)\over 2Q\delta},
\eeq

\beq
q_1^-={(2 Q^2-s_1)(2 Q^2 s-s_1 s_2)\over 4Q^3\delta}.
\eeq
%%%%%%%%%%%%
As discussed in section 3 , a crucial step in the derivation of the sum rule is
the appropriate choice of the projector $n^{\lambda}$.
Its real part will be denoted as
\beq
n_R= (n^+,n^-, {\alpha q_\perp\over \sqrt{-P^2}})
\eeq
We easily get the expressions
\beq
\alpha={(4 Q^4 - 4 Q^2 s + s_1 s_2)\over 2 Q^2}
\eeq
\beq
P^2={(2 Q^2 -s)(2 Q^2 - s_1)(2 Q^2 -s_2)(2 Q^2 s - s_1 s_2)\over 4 Q^4}
\eeq
Note that $P^2, \alpha <0$.

The relation
\beqa
n\cdot p_1 &=&\left(1/\sqrt{-P^2}\right) (n^+ p_1^- + n_- p_1^+) \\ \nonumber
 &=& {(s- 2 Q^2)(-2 Q^2 s + s_1 s_2)\over 2 Q^2 },
\eeqa
together with the condition $n\cdot p_1=n\cdot p_2$
give
\beq
{1\over n\cdot p_1 n\cdot p_2}= {(2 Q^2-s_1)(2 Q^2-s_2)\over (2 Q^2-s)(2 Q^2 s
- s_1
s_2)}.
\eeq
At the pion pole ($s_1=s_2=0, Q^2=-t/2$) we get
\beq
{1\over n\cdot p_1 n\cdot p_2}= {-t\over s (s+t)},
\eeq
which is the result quoted in
eq.~(\ref{pole}).

\renewcommand{\theequation}{B.\arabic{equation}}
\setcounter{equation}{0}
\vskip 1cm \noindent
\noindent {\large\bf Appendix B. Gluonic corrections.}
\vskip 3mm \noindent

%section{Appendix. Gluonic corrections}

The contribution to the structure $\Gamma^{(12)}$ for the diagram in Fig. 3b
can be obtained from the expression
\beqa
&&\Gamma^{12}_{3b}=(-1/4) G_{\alpha\beta}G_{\rho\sigma}  \nonumber \\
&&\times {i\over (2 \pi)^4}\int d^4\, k \,{Tr\left(
\not{n}\gamma^{\alpha}\not{k} \gamma^{\beta}\not{n}\gamma^{\rho}
 (\not{p_2}- \not{k})\gamma^{\sigma}\gamma^{\lambda}(\not{p_1}-\not{k}
+\not{q_1}) \gamma^{\lambda}(\not{p_1}-\not{k})\right)\over (p_1-k)^2 (k^2)^2
\left((p_2
-k)^2\right)^2 (p_1-k + q_1)^2}\nonumber \\
&&
\label{loop}
\eeqa
whose evaluation proceeds as follows.
Following our discussion of section 6 we approximate
eq.~(\ref{loop}) with the double dispersive contribution
\beq
\label{disp}
\Gamma^{12}_{3b}=-{1\over 4
\pi^2}\int_{0}^{\lambda^2}ds_1\int_{0}^{\lambda^2}ds_2\,
{\Delta^{3b}(s_1,s_2,s,t) \over (s_1 - p_1^2)(s_2-p_2^2)}
\eeq

Introducing the notation
\beqa
&& I[m_1,m_2,f(k^2,k\cdot p_1,...)]  \nonumber \\
\eeqa
\beq
 =\int d^4k \,{ f(k^2,k\cdot p_1,...)\delta_+(k^2-m_1^2)\delta_+((p_1-k)^2)
\delta_+((p_2-k)^2 - m_2^2)\over (p_1-k + q_1)^2}  ,
\eeq
%\\
%\beq
%I'[m_1,m_2,f(k^2,k\cdot
%p_1,...)]=\int d^4k
%f(k^2,k\cdot p_1,...)\delta_+(k^2- m_1^2)\delta_+((p_1-k)^2)\delta_+((p_2-k)^2
%%- m_2^2), \eeq
one can give a very compact form to each discontinuity integral.
After averaging over the gluon field, the spectral function associated to
eq.~(\ref{loop}),  in full analogy with eq.~(\ref{disc2}) is
given by
 \beq
\Delta(s_1,s_2,s,t)=\left({\partial\over \partial m_1^2}{\partial\over
 \partial m_2^2}I[m_1,m_2,f(k^2,k\cdot p_1,...)]\right)_{m1,m_2=0}
\label{de}
\eeq
with
\beqa
f(k^2,k\cdot p_1,...)&=& 64 (k\cdot n -n\cdot p_1) \nonumber \\
                     & & \times \, (k^2 k\cdot n -2 k\cdot n k\cdot p_1
-2 k\cdot n k\cdot q_1 \nonumber \\
  & & \quad \quad + k\cdot q_1 n\cdot p_1 + k\cdot n s_1 -k\cdot n p_1\cdot
q_1).
\label{effe}
\eeqa
where we have used the relations
\beq
n\cdot q_1=0,\,\,n\cdot n=0,\,\,\,\, n\cdot p_1=n\cdot p_2.
\eeq

Notice that the leading asymptotic behaviour  in eq.~(\ref{effe}) comes from
the term containing
the factor $ n\cdot k \,n\cdot p_1,\, (n\cdot p_1)^2$ and
$ n\cdot p_1\, n\cdot k\,k\cdot q_1$ which give an asymptotic spectral
function of the form
\beqa
&&\Delta^{(12),asymp}_{pc,3a} \nonumber \\
&& ={5\over 27} \langle G^2 \rangle
\{ I[n\cdot p_1\, n\cdot k]\left(
k^2 -2 k\cdot p_1 + s_1 \right. +I[(n\cdot p_1)^2 k\cdot q_1]\nonumber  \\
 & & \quad \quad \quad-3I[n\cdot p_1\, n\cdot k\, k\cdot q_1] \}.
\label{asy}
\eeqa
Only the leading and first subleading contributions in power of $Q$, the
momentum in the
"plus" light-cone direction of the incoming pion, are kept.
The evaluation of the integrals in eq.~(\ref{asy}) is rather lenghthy. Again,
the method to
use in this case is very similar to the one followed for the scalar case.
For completeness we give here the results of the evaluation of 3 main integrals
which appear in the case of diagram $3a$
\beqa
Y_1 &=& I[n\cdot p_1\, n\cdot k] \nonumber \\
    &=& {4 \pi Q^4 (-2 Q^2 + s)(-s + s_1)(-2 Q^2 s + s_1 s_2)\over s^3 (2 Q^2 -
s_1)^3
(-2 Q^2 + s_2)^3};
\eeqa

\beqa
Y_2 &=& I[(n\cdot p_1)^2\, n\cdot k] \nonumber \\
   &=& {2 \pi Q^4 (-2 Q^2 +s )(-s + s_1)^2(-2 Q^2 s + s_1 s_2)\over
s^3 (-2 Q^2 + s_1)^3 (-2 Q^2 + s_2)^3};
\label{Y1}
\eeqa
We also find
\beqa
Y_3 &=&I[n\cdot p_1\, n\cdot k\, k\cdot q_1] \nonumber \\
     &=& Y_2.
 \label{Y3}
 \eeqa
A simple calculation shows that $Y_1\approx 1/Q^6$, while $Y_2,Y_3 \approx
1/Q^4$.
Other integrals which appear in the power corrections to the spectral function,
such as those containing additional powers of $n\cdot k$ are suppressed even
more
by additional powers of $Q$.
Notice also that all integrals we have considered have been evaluated with the
same method
used to derive eq.~(\ref{disc3}).
The diagrams in Figs. 3b and 3.c can be evaluated similarly.
For 3.b we shift with mass parameters the quartic propagators
obtained from the insertion of the external gluon line and proceed as in
eqs.~(\ref{disc1},\ref{disc2},\ref{disc3}). Those terms proportional to
additional powers of
$n\cdot k$ can also be evaluated by this method. Their contribution has been
included
only in the lowest order perturbative spectral density
eqs.~(\ref{depert12}). In the case of the power corrections such additional
contributions are not going to affect the result to leading and first
subleading power in
$1/Q$.

\newpage

\noindent
{\bf Figure Captions}
\bigskip

 \bigskip

 \noindent
 1. The correlator at lowest order in $\alpha_s$.
\smallskip

\noindent
2a, 2b. Typical quark power corrections.
\smallskip

 \noindent
3a, 3b, 3c. gluonic power corrections.
\smallskip

\noindent
The integration contours for the scalar amplitude in the $p_1^2$ plane, showing
the $u=0$ threshold.
The shadowed region denotes the area enclosed by the modified Borel transform.
\smallskip

\noindent
5a. Lowest order perturbative contribution to the spectral density for pion
Compton scattering.
The dashed lines are put on shell.
\smallskip

\noindent
5b. Lowest order perturbative contribution to the spectral density of the pion
form factor.
\smallskip

 \end{document}